\documentclass[aps,prd,preprint,groupedaddress,showpacs,showkeys]{revtex4}
\begin{document}
\title{Semiclassical (QFT) and Quantum (String) anti - de Sitter Regimes: New Results}
\author{A. Bouchareb$^{1,2}$, M. Ram\'on Medrano$^{3,2}$ and N. G. S\'anchez$^{2}$}
\email{norma.sanchez@obspm.fr, mrm@fis.ucm.es}
 
\affiliation{
(1) Departament de Physique, Universit\'e d' Annaba, B.P.12, El-Hadjar, Annaba 23000, Algeria.\\
(2) Observatoire de Paris, LERMA, CNRS UMR 8112,  61, Avenue de l'Observatoire, 75014 Paris, France. \\
(3) Departamento de F\'{\i}sica Te\'orica, Facultad de Ciencias F\'{\i}sicas, Universidad Complutense, 
E-28040 Madrid, Spain}

\date{\today}
\begin{abstract}
We compute the quantum string entropy $S_s(m, H)$ from the microscopic string density of states $\rho_s (m,H)$ of mass $m$ in Anti de Sitter space-time. For high $m$, (high $Hm \rightarrow  c/\alpha' $), no phase transition occurs at the Anti de Sitter string temperature $ T_{s}= (1/2\pi k_B)L_{c\ell}~c^2/\alpha'$, which is higher than the flat space (Hagedorn) temperature $t_{s}$. ($L_{c\ell}= c/H$,  the Hubble constant $H$ acts as producing a smaller string constant $\alpha'$ and thus, a higher tension). $ T_s$ is the precise quantum dual of the semiclassical (QFT) Anti de Sitter temperature scale $T_{sem}=\hbar c /(2\pi k_B L_{c\ell})$.
We compute the quantum string emission $\sigma_{string}$ by a black hole in Anti de Sitter (or asymptotically Anti de Sitter) space-time (bhAdS).  For $T_{sem~bhAdS}\ll T_{s}$, (early evaporation stage), it shows the QFT Hawking emission with temperature $T_{sem~bhAdS}$, (semiclassical regime). For $T_{sem~bhAdS}\rightarrow T_{s}$, it exhibits a phase transition   into a  Anti de Sitter string  state of size $L_s = \ell_s^2/L_{c\ell}$, ($\ell_s= \sqrt{\hbar \alpha'/c}$), and Anti de Sitter string temperature $T_s$. New string bounds on the black hole emerge in the $bhAdS$ string regime.
The $bhAdS$ string regime determines a maximal value for $H$: $H_{max}= 0.841 c/l_s$. The minimal black hole radius in $AdS$ space time turns out to be $r_{gmin}= 0.841l_s$, and is larger than the minimal black hole radius in $dS$ space time by a numerical factor equal to 2.304. We find a  new formula for the full Anti de Sitter entropy $S_{sem} (H)$, as a function of the usual Bekenstein-Hawking entropy $S_{sem}^{(0)}(H)$. For $L_{c\ell}\gg \ell_{Planck}$ , ie. for low $H\ll c/\ell_{Planck}$, or classical regime,  $S_{sem}^{(0)}(H)$ is the leading term with its logaritmic correction, but for high $H \geq c/\ell_{Planck}$ or quantum regime,  no phase transition operates, in contrast to de Sitter space, and the entropy $S_{sem} (H)$ is very different from the Bekenstein-Hawking term $S_{sem}^{(0)}(H)$

\end{abstract}
\pacs{}
\keywords{ Anti de Sitter background, semiclassical gravity, quantum gravity, quantum strings, black-hole - Anti de Sitter background, classical/quantum duality}
\maketitle
\section{Introduction and Results\label{sec:intro}}

The understanding of the semiclassical and quantum regimes of gravity in the contexts of quantum field theory and string theory respectively is particularly important for several reasons and has been the object of recent intensive study with two relevant examples: rotating black holes and de Sitter background in both regimes~\cite{1}~\cite{2}. Naturally in this context, one regime is the classical/quantum  gravity dual of the other, in the precise sense of the usual classical/quantum (wave-particle) duality~\cite{3}-\cite{6}. 

In this paper, we describe classical, semiclassical and quantum Anti de Sitter ($AdS$) regimes. A clear picture for Anti de Sitter background is emerging, going beyond the current picture, both for its semiclassical and quantum regimes. Although there is no relevant cosmological motivation to consider Anti de Sitter background, the understanding of semiclassical and quantum $AdS$ regimes is particularly important for the following reasons:
(i) $AdS$ regimes are illustrative examples to compare 
and contrast with de Sitter regimes, and to see the effects of a negative cosmological constant;
(ii) $AdS$ space time provides asymptotic boundary conditions to Black Hole evaporation and also gives a natural infrared cuttoff to the euclidean path integral formulation for quantum gravity, $AdS$ space time acting as a large space box for it~~\cite{7}-\cite{9};
(iii) the results of classical and quantum string dynamics in Anti de Sitter space-time: Solving the classical and quantum string dynamics in conformal and non conformal invariant string backgrounds it shows that the physics is the same in the two class of backgrounds: conformal and non conformal. The mathematics is simpler in conformal invariant backgrounds (WZWN models), and the main physics, in particular the string mass spectrum, remains the same \cite {10}-\cite{14}.

For strings in  flat space-time, weak string coupling  does not describe Anti de Sitter (AdS) background , or black holes in  AdS background, but for strings in the full curved  AdS background or in the black hole AdS background, as we consider here, AdS and the black hole are non-perturbative from the beginning.  In the context of gravity, for the  thermodynamical behavior of black holes or for the thermodynamics of strings in curved background, string-string  interactions do not seem be crucial. We start from the beginning with the full non perturbative AdS or AdS-black hole background, so strong coupling effects are present even if no explicit string self interaction is added. Of course, more interactions can be included and explored, but strings in curved background provide at least a framework to deal with strong gravity regimes. 

A central object in string theory is the string density of states, in particular its high mass behavior, which depends on the different curved space-times in which strings propagate (this is the Hagedorn behavior in flat space time). For a given space-time, this mass behavior grows exponentially; aspects as the number of space-time dimensions, critical dimensions, type of strings, type of string theory only appear in the two dimensionless numerical coefficients of the exponential growth and its amplitude.  This is precisely the importance of the density of string states, from which the string intrinsic temperature and entropy emerge, and from which, the basic important results of string thermodynamics, are derived.

An important object of our study is $\rho_s (m,H)$, the microscopic string density of states of mass $m$ in $AdS$ background. $\rho_s (m,H)$ is derived from the string density of levels $d(n)$ of level $n$ and from the string mass spectrum $m (n, H)$ in $AdS$  background. The mass formula $m (n, H)$ is obtained by solving the quantum string dynamics in $AdS$ space time \cite {11} \cite{13} \cite{14}. The density of levels $d(n)$ is the same in flat and in curved space-times. As a result, the mass formula $m (n, H)$ and $\rho_s (m, H)$ are different from the respective flat space-time string mass spectrum and flat space mass level density. The formulae $m(n, H)$ and $\rho_s (m, H)$ depend on the two characteristic lengths in the problem: the classical length $L_{c\ell} = c/H$, and the fundamental string length $\ell_s =  \sqrt{\hbar \alpha'/c}$, or equivalently on the respective mass scales: $M_{c\ell} = c^2 L_{c\ell}/G$ (D=4) and $ m_s =\ell_s / \alpha'$. Relevant combinations of them emerge in the mass formula $m(n,H)$ as the Anti de Sitter string length $L_s =\hbar/c M_s$ or Anti de Sitter string mass $M_s = L_{c\ell}/\alpha'$. Contrary to strings in de Sitter background, the temperature $T_s=M_sc^2/2\pi k_B$ is not a critical string temperature for strings in Anti de Sitter background. The mass spectrum and the density of mass levels of quantum string states in $AdS$ background do not exhibit any upper mass bound as in $dS$ background.

From the string density of states, $\rho_s (m, H)$, we obtain the string entropy $S_s(m, H)$. We find that no phase transition takes place at $m = M_s$,~ ie~  $T = T_s$, in contrast to   strings in flat space time, de Sitter and black hole backgrounds. The entropy $S_s(m, H)$ of string states in $AdS$ space time is larger than the string entropy for $H=0$. The effect of a negative cosmological constant is to increase the entropy, while in $dS$ space time the effect of a positive cosmological constant is to reduce the entropy.

The $AdS$ string temperature $ T_{s}=  \Big(L_s/\ell_s\Big) t_s $ is higher than the flat space (Hagedorn) value $t_s $. This is so, since for high masses, the string temperature in $AdS$ background is $T_s$, instead of $t_s$. The flat space (Hagedorn) temperature $t_s$ is the scale temperature in the low $Hm$ regime. The constant $H$ acts in the temperature as an "effective string tension" $(\alpha^{'}_H) ^{-1} $ : a smaller  $\alpha^{'}_H = (\hbar/c)\Big(~H \alpha'/c\Big)^2$, (and thus a higher tension). $T_{s}$ turns out to be the precise quantum dual of the semiclassical (QFT) Anti de Sitter temperature scale $T_{sem}=\hbar c /(2\pi k_B L_{c\ell})$, the two temperatures satisfy: $T_{s} = t_{s}^2~ T_{sem}^{-1}$. 

There is no critical string temperature in $AdS$ background, while there exists a critical temperature in $dS$ background, as well as in the black hole backgrounds. In $AdS$ background, the density of states and the entropy do not show any singular behavior at finite mass or temperature. As a consequence, there is  no string phase transition of the Hagedorn type in $AdS$ background.

For low $Hm\ll c/ \alpha'$, the partition function of a string gaz in $AdS$ background  shows a pole singularity at $T_{sem} \rightarrow t_s$. This transition, near the flat space string temperature $t_s$ and for any dimension $D$, is the same as the string transition in flat space time (Carlitz/Hagedorn transition). This is also the same as the string transition far from the black hole in the Schwarszchild black hole space time, and as the string transition in the low curvature regime ($mH \ll c/\alpha'$) of $dS$ background.
The partition function for excited and highly excited strings in $AdS$ background does not feature any singular behavior at $T_s$, in contrast to the string partition function in flat space time which shows a single pole temperature singularity (Carlitz transition), and also in contrast to the string partition function in $dS$ background which shows a square root branch point singularity (de Vega-Sanchez transition)~\cite{15}. There is no string phase transition at $T_s$ for massive and highly massive strings in $AdS$ alone, that is for $Hm\alpha'/c \sim 1$ and for $Hm\alpha'/c \gg 1$, in contrast to strings in $dS$ background.

We also consider the string regimes of a black hole in Anti de Sitter (or asymptotically) Anti de Sitter background ($bhAdS$). We compute the quantum string emission cross section $\sigma_{string}$ by a  Schwarschild black hole in Anti de Sitter background. 
For $T_{sem~bhAdS}\ll T_{s}$, (early evaporation stage), $\sigma_{string}$ shows the Hawking emission with temperature $T_{sem~bhAdS}$, (semiclassical regime). For $T_{sem~bhAdS}\rightarrow T_{s}$, $\sigma_{string}$ exhibits a phase transition at $T_{sem~bhAdS} = T_{s}$: the  massive emission condensates into an $AdS$ string state of Anti de Sitter string size $L_s$ and $AdS$ string temperature $T_s$. 

For $m\ll M_s $, (low regime $Hm \ll c/\alpha'$) the $bhAdS$ string emission cross section shows the same singular behavior near $t_s$ as the low $Hm$ behavior of the $AdS$ partition function, and as the quantum string emission by an asymptotically flat black hole, but here at the $bhAdS$ temperature. This is so, since in the $bhAdS$ background the string mass scale for low string masses (temperatures) is the Hagedorn flat space time temperature $t_s$. 

The evaporation of a black hole in $AdS$ space time, from its early semiclassical or quantum field theory regime (Hawking radiation) towards a quantum string $AdS$ regime (late stages), is seen as well from the black hole decay rate. In the early evaporation stages, the semiclassical black hole in $AdS$ background decays thermally as a grey body at the Hawking temperature $T_{sem~bhAdS}$ with the decay rate $\Gamma_{sem}\sim G (T_{sem~bhAdS})^{3}$ . As evaporation proceeds $T_{sem~bhAdS}$ increases until it reaches the string $AdS$ temperature $T_s$, the black hole itself becomes an excited string $AdS$ state, decaying with a string width $\Gamma_s \sim \alpha' T^{3}_s$ into all kind of particles, with pure (non mixed) quantum radiation. For $T_{sem~bhAdS} \rightarrow T_s$, the semiclassical black hole decay rate becomes precisely the quantum string decay $\Gamma_s$.

New string bounds on the black hole emerge in the $bhAdS$ string regime, i.e. when $T_{sem bhAdS}= T_s$. The $bhAdS$ space time allows an intermediate string regime, not present in the Schwarzschild black hole alone ($H=0$): in an asymptotically flat space time, the black hole radius $r_g$ becomes $l_s$ in the string regime;
in an asymptotically $AdS$ space time, when $T_{sem~bhAdS}=T_s$, the black hole radius can have two different values depending on the $AdS$ string size.
There is a minimal $AdS$ length $L_{c\ell}=1.189~~l_s $. 
The $bhAdS$ string regime determines a maximal value for $H$: $H_{max}=0.841 c/l_s$.
The minimal black hole radius in $AdS$ space time turns out to be $r_{g min}= 0.841~~ l_s$, and is larger than the minimal black hole radius in $dS$ space time by a numerical factor equal to $2.304$.                        

By precisely identifying the semiclassical and quantum (string) Anti de Sitter regimes, we find a new formula for the Anti de Sitter entropy $S_{sem} (H)$ with quantum corrections included, and which is a function of the Bekenstein-Hawking entropy $S_{sem}^{(0)}(H)$.
The classical $AdS$ regime is like the classical $dS$ regime, or the classical Schwarschild black hole regime ($M_{c\ell}\gg m_{Pl}$), that is in this regime the leading term to the entropy is the Bekenstein-Hawking entropy. But for a large Hubble constant ($M_{c\ell}\sim m_{Pl}$), $S_{sem}^{(0)}(H)$ is subdominant and the whole entropy $S_{sem}(H)$ is different from the Bekenstein-Hawking entropy $S_{sem}^{(0)}(H)$.
Contrary to de Sitter space, no phase transition occurs at $T \rightarrow T_{sem}$ in $AdS$ space time.  That is, there is no phase transition at $T \rightarrow t_{Pl}$, in the high curvature $H \rightarrow c/\ell_{Pl}$ or quantum $AdS$ regime.  This is so, since for AdS background,
like for the AdS string entropy, no singularity at finite mass or finite temperature occurs in the density of states or in the entropy.
Also contrary to de Sitter space, and to the rotating black hole (extremal transition), there is not a square root branch point behavior in the mass (temperature) analogous to the thermal self-gravitating gas phase transition of point particles \cite {15}.

\section{Semiclassical  Anti de Sitter background\label{sec:semiclassical}}

In static coordinates, the D-dimentional anti de Sitter (AdS) space-time is described by the metric 
\begin{equation}
ds^{2}=-B(r) ~c^2~dt^{2} + B^{-1}(r) ~dr^2 + r^2 ~d\Omega_{D-2}^2
 \label{eq:m1}
\end{equation}
where: 
\begin{equation}
B(r) = 1 + \frac{H^2~r^2}{c^2}
 \label{eq:m2}
\end{equation}

AdS space time is generated by a negative cosmological constant $\Lambda$, or a constant positive pressure density $p$ satisfying the equation of state $\epsilon~+~p=0$, with $\epsilon$ being a constant negative density of energy.

In terms of $H^2 ~(H^{2} > 0)$, the cosmological constant $\Lambda$, and the  scalar curvature $R$ are given by 
\begin{equation}
\Lambda = - \frac{(D-2)(D-1)}{2} ~\frac{H^2}{c^2}
\label{eq:lambda}
\end{equation}

\begin{equation}
R = - D(D-1) \frac {H^2}{c^2} 
\label{eq:H}
\end{equation}

Notice that one can go from de Sitter $(dS)$ to anti de Sitter $(AdS)$ background, through the change $H \rightarrow iH$. 

The characteristic AdS length, mass and temperature associated to $| \Lambda |$ , or $|H|$ are given by:

\begin{equation}
L_{c\ell} = \sqrt{\frac{(D-2)(D-1)}{2~|\Lambda |}} = \frac{c}{|H|}
\label{eq:Lcl}
\end{equation}
\begin{equation}
M_{c\ell} = \frac{c^{2}}{G} ~L_{c\ell}=\frac{c^{3}}{G~|H|} ~~~~~~~~~\text{(D=4)}
\label{eq:Mcl}
\end{equation}
\begin{equation}
T_{sem} = \frac{\hbar ~c}{2\pi~k_{B}} ~ \frac{1}{L_{c\ell}}=\frac{\hbar }{2\pi~k_{B}} ~ |H|
\label{eq:Tcl}
\end{equation}

From Eqs.~(\ref{eq:m1}) and (\ref{eq:m2}), it is clear that there is no event horizon in anti de Sitter space. Nevertheless, Eq.(\ref{eq:Tcl}) defines the semiclassical AdS temperature.  This a measure (in units of temperature) of the semiclassical mass scale:
\begin{equation}
M_{sem}=\frac{m_{pl}^2}{M_{c\ell}}
\label{eq:Msem}
\end{equation}

Eq.(\ref{eq:Tcl}) is for AdS the analogous of the Hawking temperature.

Analogously to de Sitter space-time, we can write the (zeroth order) AdS  semiclassical entropy $S_{sem}^{(0)}(H)$ (from now on we write $|H|\equiv H$):
\begin{equation}
S_{sem}^{(0)}(H) =  \pi k_B ~ \Bigg(  \frac{L_{c\ell}}{\ell_{pl}}\Bigg)^2 =\pi k_B ~ \Bigg(\frac{M_{cl}}{m_{pl}}\Bigg)^2
\label{eq:SH0}
\end{equation}
where
\begin{equation}
\ell_{pl} =  \sqrt{\frac{\hbar ~ G}{c^3}}~~;~~~~m_{pl}=\sqrt{\frac{\hbar ~ c}{G}}
\label{eq:lmPl}
\end{equation}
$\ell_{pl}$ and $m_{pl}$ being the Planck length and Planck mass respectively,  and $M_{cl}$ is the (classical) mass scale of $AdS$ background Eq. (\ref{eq:Mcl}).

Eq.~(\ref{eq:SH0}) is better expressed as
\begin{equation}
S_{sem}^{(0)}(H) = \frac{1}{2} ~\frac{M_{cl}~ c^2}{T_{sem}}
\label{eq:S0cl}
\end{equation}
where $T_{sem}$ is the semiclassical  AdS temperature defined  by Eq.(\ref{eq:Tcl}). In terms of the classical and the semiclassical masses, $T_{sem}$ is expressed as 
\begin{equation}
T_{sem} =  \frac{c^2}{2\pi k_B} ~\frac{m_{pl}^2}{M_{c\ell}}=\frac{1}{2\pi k_B}~M_{sem}~c^2
\label{eq:TM}
\end{equation} 

Eq.~(\ref{eq:SH0})or (\ref{eq:S0cl}) is the Bekenstein-Hawking entropy for AdS background~~\cite{16}~~\cite{17}. It is the ordinary entropy expression for any ordinary system, where $T_{sem}$ is the semiclassical temperature. $T_{sem}$ is the Compton length of AdS space in the units of temperature, that is, the temperature scale of the semiclassical gravity properties for which the {\it mass scale} is precisely $M_{sem}$ Eq.~(\ref{eq:Msem}). This semiclassical or intermediate energy regime interpolates between the classical and the quantum regimes of gravity. We discuss more on these regimes in Section~(\ref{sec:dual}). 

The full  semiclassical entropy $S_{sem}(H) $ is related to the semiclassical density of states $\rho_{sem}(H)$ through the relation
\begin{equation}
\rho_{sem}(| \Lambda |)~= ~e^{\frac{S_{sem} (| \Lambda |)}{k_{B}}}
\label{eq:Scl}
\end{equation}

The full semiclassical AdS entropy will be consider in Section ~(\ref{sec:dual}).

\section{Quantum String Entropy in Anti de Sitter Background \label{sec:qs}}

The full string entropy $S_s(| \Lambda |)$ is related to the microscopic string  density of states $\rho_{s}(| \Lambda |)$ by
\begin{equation}
\rho_{s}(| \Lambda |)~= ~e^{\frac{S_s (| \Lambda |)}{k_{B}}}
\label{eq:rhos}
\end{equation}

For high n, the mass formula for quantum strings in an anti de Sitter background is ~\cite{11}~~\cite{13}~~\cite{14}

\begin{equation}
 \frac{m}{m_{s}} \simeq  n  \left(\frac{1}{n} + \left(\frac{m_{s}}{M_{s}} \right)^{2} \right)^{\frac{1}{2}} ~~~~~~~~~~~~~~~~~ \text{open strings}
\label{eq:spo}
\end{equation}

\begin{equation}
 \frac{m}{m_s} \simeq  2~n  \Bigg(\frac{1}{n} + \Bigg(\frac{m_s}{M_s} \Bigg)^2 \Bigg)^{\frac{1}{2}} ~~~~~~~~~~~~~~~~~\text{closed strings}
\label{eq:spc}
\end{equation}

$m_{s}$ and $M_s$ are given by 

\begin{equation}
m_{s}=\sqrt{\frac{\hbar}{\alpha' c}} = \frac{\ell_s}{\alpha'}~~~;~~~M_{s}= \frac{L_{c\ell}}{\alpha'} = \frac{c}{\alpha'H}~~~,~~~\frac{m_s}{M_s}=\frac{\ell_s}{L_{c\ell}}=\frac{H}{c}\ell_s
\label{eq:mMs}
\end{equation}

$\alpha'$ is the fundamental string constant, ($\alpha'^{-1}$ is a linear mass density), $m_s$ and $\ell_s$ are the fundamental (flat space) string mass and length respectively, $L_{c\ell}$ is given by Eq.~(\ref{eq:Lcl}) and $M_{s}$ is the characteristic string mass in AdS.  
Furthermore, $M_s$ defines the quantum string Anti de Sitter length $L_s$:
\begin{equation}
L_s = \frac{\hbar }{M_s ~c}= \frac{\ell_s^2}{L_{c\ell}}=\frac{\hbar\alpha'}{c^2}H~~,
\label{eq:Ls}
\end{equation}
and the string AdS temperature $T_s$:
\begin{equation}
T_s = \frac{1}{2\pi k_B} ~ M_s~c^2 = \frac{\hbar c}{2 \pi k_B} ~\frac{1}{L_s}= \frac{1}{2 \pi k_B} ~\frac{c^3}{H \alpha'} 
\label{eq:Ts}
\end{equation}
As we will see in in  Sections ~\ref{sec:sdsps}, \ref{sec:partition} and ~\ref{sec:bBHS}, $T_s$ plays a relevant role as a limiting temperature.

For higher mass states (very large $n$), the mass spectrum behavior is
\begin{equation}
m \simeq  n ~ \left(\frac{m_{s}^{2}}{M_{s}} \right)= \frac{n}{\alpha'} ~ \left(\frac{\ell_{s}^{2}}{L_{c\ell}}\right)= n ~\Big(\frac{\hbar}{c}\Big) \Big(\frac{H}{c}\Big)
\label{eq:nMs} 
\end{equation}
which differs drastically from the flat space time string behavior, 
as well as from the de Sitter high mass string behaviour  $m ~\simeq ~ m_s \sqrt{n}$. 

Notice that the ratio  $(m_{s}^{2}/ M_{s})$  does   {\bf not} depend on the string constant $\alpha'$ (Eqs.~(\ref{eq:mMs}), (\ref{eq:Lcl})):
\begin{equation}
 m \sim  n ~ \frac{\hbar}{c ~L_{c\ell}} 
\label{eq:nLcl} 
\end{equation}
Thus, for high $m$, the characteristic anti de Sitter length contains $n$ times the string Compton wave length: $L_{c\ell} \sim n ~(\hbar/cm)$.  
 
From Eqs.~(\ref{eq:spo}) and (\ref{eq:spc}), in the limit $| \Lambda | \rightarrow 0$ (see Eqs.~(\ref{eq:Lcl}) and (\ref{eq:mMs})), we recover the flat space time string spectrum:
\begin{equation}
m^2 \simeq ~m_{s}^2~ n ~~~~~\text{(open)}, ~~~~~~m^2 \simeq ~m_{s}^2~ 4 n~~~~~~\text{(closed)}
\label{eq:lam01} 
\end{equation}

In order to derive $\rho_{s}(m, | \Lambda |)$, let us notice that the degeneracy $d_{n}(n)$ of level $n$ (counting of oscillator states) is the same in flat and in curved space time. The differences, due to background curvature, enter through the relation $m=m(n,H)$ of the mass spectrum. As is known, asymptotically for high $n$, the degeneracy $d_n(n)$ behaves universally as 

\begin{equation}
d_{n}(n) = n^{-a'}~ e^{b~ \sqrt{n}}
\label{eq:d}
\end{equation}
where the constants $a'$ and $b$ depend on the space time dimensions and on the type of strings; for bosonic strings :
\begin{equation}
b=2~\pi\sqrt{\frac{D-2}{6}}~;~~~~~a'=\frac{D+1}{4}~~~\text{(open)}~;~~~~~~a'=\frac{D+1}{2}~~~\text{(closed)}
\label{eq:ba'}
\end{equation}

Let us observe that the density $\rho_{s}(m, | \Lambda |)$ of mass levels and the level degeneracy $d_{n}(n)$ satisfy
\begin{equation}
\rho_s(m, |\Lambda|) ~d \left(\frac{m}{m_{s}}\right)= d_{n}(n) ~dn 
\label{eq:rod1}
\end{equation}
ie.
\begin{equation}
\rho_{s}(m,|\Lambda |) \simeq \frac{m}{m_s} \left[ \frac{ d_{n}(n)}{g^{'}(n)} \right]_{n=n(m)} 
\label{eq:rod2}
\end{equation}
where $\left(m /m_{s}\right)^{2} \simeq g(n)$; $g(n)$ is read directely from the r.h.s of Eqs.~(\ref{eq:spo}) and~(\ref{eq:spc}).

We obtain then, for open strings, the following density of mass levels:
\begin{equation}
 \rho_s( m, |\Lambda |) \simeq  \frac{m}{m_s}~ \Bigg[ \frac{-1+\sqrt{1+4(\frac{m}{M_s})^2}}{2(\frac{m_s}{M_s})^2}~\Bigg ]^{-a'}~~
 \frac{\exp \Bigg\{b~\Bigg[ \frac{-1+\sqrt{1+4(\frac{m}{M_s})^2}}{2(\frac{m_s}{M_s})^2}~ \Bigg ]^{1/2} \Bigg\}}
{\sqrt{1+4(\frac{m}{M_s})^2}}
 \label{eq:rhoo} 
\end{equation}
Analogously, for closed strings:
\begin{equation}
\rho_s( m, |\Lambda |) \simeq  \frac{m}{m_s}~ 
\Bigg[~ \frac{-1+\sqrt{1+(\frac{m}{M_s})^2}}{2(\frac{m_s}{M_s})^2}~\Bigg ]^{-a'}~
\frac{\exp \Bigg\{b~\Bigg[ \frac{-1+\sqrt{1+(\frac{m}{M_s})^2}}{2(\frac{m_s}
{M_s})^2}~ \Bigg ]^{1/2} \Bigg\}}{\sqrt{1+(\frac{m}{M_s})^2}}
\label{eq:rhoc} 
\end{equation}
$\rho_s$ depends on $|\Lambda |$ through $M_s$. Notice that in anti de Sitter background, $\rho_s$ does not exhibit any branch point singularity as in de Sitter case ~\cite{2}. 

For $| \Lambda | \rightarrow 0$, we recover the flat space time string solutions:\\
\begin{equation}
 \rho_{s} \Big(\frac{m}{m_s}\Big) \simeq  \Big( \frac{m}{m_s} \Big)^{-a} ~e^{\frac{b}{2} \big( \frac{m}{m_s} \big)} ~~~~~~\text{(closed)}
\label{eq:rhofc}
\end{equation}
and \\
\begin{equation}
\rho_s(m) \simeq  \Bigg( \frac{m}{m_s} \Bigg)^{-a} ~e^{b \big( \frac{m}{m_s} \big)} ~~~~~~\text{(open)}
\label{eq:rhofo}
\end{equation}
where $a~=~2a'~-~1$. 

Several expressions for the exact $\rho_s(m,H)$ are useful depending on the different behaviors we would like to highlight: the  flat H=0 limit, the low mass, or the high mass behavior.\

$\rho_s(m,H)$ ,Eq.~(\ref{eq:rhoo}) or (\ref{eq:rhoc}), can be expressed in a more compact way as:

\begin{equation}
\rho_s (m, H)\simeq \Bigg (\frac{m}{\Delta_s M_s}\sqrt{\frac {2}{\Delta_s - 1 }}\Bigg)~ 
\Bigg(\frac{M_s}{m_s}\sqrt{\frac{\Delta_s - 1 }{2}}\Bigg)^{-a}~ e^{\Big(     \frac{bM_s}{m_s}\sqrt{\frac{\Delta_s - 1 }{2}}\Big)} 
\label{eq:rhodm}
\end{equation}
where
\begin{equation}
\Delta_s \equiv\sqrt{1 + \Big(\frac{m}{M_s}\Big)^2}
\label{eq:deltasm}
\end{equation}

Let us introduce the (zeroth order) string entropy $S^{(0)}_{s}$ in flat space time : 
\begin{equation}
S^{(0)}_{s} =   \frac{1}{2}~b~k_B \Big(\frac{m}{m_s}\Big) =\frac{1}{2}~\frac{m~c^2}{t_s} ~~~\text{(closed)}
\label{eq:S0c}
\end{equation}

\begin{equation}
S^{(0)}_{s}=   b k_B~\Big(\frac{m}{m_s}\Big) = \frac{m~c^2}{t_s} ~~~\text{(open)}
\label{eq:S0o}
\end{equation}
where
\begin{equation}
t_{s} =\frac{1}{b ~k_B}~ m_s~ c^{2}
\label{eq:ts}
\end{equation}
being $t_{s}$ the flat space string temperature.

Therefore, from Eqs.~(\ref{eq:S0c}),(\ref{eq:S0o}),(\ref{eq:rhoo})and (\ref{eq:rhoc}), the mass density of levels $\rho_s(m, H)$ for both open and closed strings can be expressed, in terms of $S^{(0)}_{s}$, as :

\begin{equation}
\rho_s (m, H) =\Big(  \frac{S_s^{(0)}}{k_B} \sqrt{f(x)} \Big)^{-a}~       e^{\Big(\frac{S_s^{(0)}}{k_B}\sqrt{f(x)}\Big)} ~~ 
\frac{1}{\sqrt{(1 + 4x^2)f(x)}}
\label{eq:rhoF}
\end{equation}

\begin{equation}
a=\frac{(D-1)}{2} ~~\text{(open)},~~  a=D ~~\text{(closed)}, ~~~~~~f(x) = \frac{ - 1 + \sqrt{1 + 4 x^2}}{2x^2}
\label{eq: FX}
\end{equation}

$x$ being the dimensionless variable
\begin{equation}
x(m, H)\equiv  \frac{1}{2}\Big(\frac{m}{M_s}\Big)= 
\frac{m_s}{b M_s}\frac{S_s^{(0)}}{k_B} 
\label{eq:X}
\end{equation}

In terms of $\Delta_s$ Eq.(\ref{eq:deltasm}), we have:

\begin{equation}
\rho_s (m, H)\simeq \frac{1}{\Delta_s}\sqrt{\frac{1 + \Delta_s}{2}}~ 
\Bigg(\frac{S_s^{(0)}}{k_B}\sqrt{\frac{2}{1 + \Delta_s}}\Bigg)^{-a}~ e^{\Big(     \frac{S_s^{(0)}}{k_B}\sqrt{\frac{2}{1 + \Delta_s}}\Big)} 
\label{eq:rhodso}
\end{equation}
 
\begin{equation}
\Delta_s \equiv\sqrt{1 + 4x^2}~~ ~~,~~~~~ f(x)= \frac{2}{1+\Delta_s}
\label{eq:deltas}
\end{equation}
For small $x$, (small $H m\alpha '/c$), $f(x)$ can be naturally expressed as a power expansion in $x$. In particular, for $H=0$, we have $x=0$ and $f(x)=1$, and we recover the flat space time string solution: 
\begin{equation}
 \rho_s(m) \simeq  \Big( \frac{S_s^{(0)}}{k_B} \Big)^{-a} ~e^{\big( \frac{S_s^{(0)}}{k_B} \big)}
\label{eq:rhoS0f}
\end{equation}
For  $ x \ll 1 $,  i.e $ m \ll M_{s} $, the corrections to the flat $(H=0)$  solution are given by: 
\begin{equation}
\rho _s(m, H)_{m\ll M_s}  \sim \left( \frac{m}
{m_s}\right) ^{-a} ~e^{\frac{b}{2}\Big(\frac{m}{m_s}\Big)\Big[~1 + ~\frac{1}{8}( \frac{\alpha'H m}{c})^2~ + ~O\left(\frac{m}{M_s}\right)^3~\Big]}
\label{eq:rlH}
\end{equation}

From Eqs.~(\ref{eq:rhos}), (\ref{eq:rhoF}) we can read the full string entropy in Anti de Sitter space :

\begin{equation}
S_s(m,H) = \hat {S_s}^{(0)}(m,H) 
-a~k_B~\ln ~\big(\frac{\hat {S_s}^{(0)}(m, H)}{k_B}\big) - k_B ~\ln F(m,H)
\label{eq:SsHF}
\end{equation}
\begin{equation}
\hat {S_s}^{(0)}(m,H)\equiv S_s^{(0)}\sqrt{f(x)}~~~~~,~~~~  F(m,H) \equiv \sqrt{(1 + 4x^2)f(x)}
\label{eq:SsdsHF}
\end{equation}
that is : \begin{equation}
S_s(m,H) = \sqrt{f(x)} ~ S_s^{(0)}  
-a~k_B~\ln \Big(\frac{\sqrt{f(x)}~S_s^{(0)}}{k_B}\Big) - k_B \ln \sqrt{f(x)}- k_B \ln~\sqrt{~1 + 4x^2~}
\label{eq:SsH}
\end{equation}

The string  mass domain in Anti de Sitter background is $ 0 \leq  m \leq \infty$, ie. $ 0 \leq x \leq \infty $,  (which implies $1\leq \Delta_s  \leq  \infty$,  ie. $0 \leq f(x)\leq 1 $). Notice that for de Sitter background, $\Delta_s = \sqrt{1 - 4x^2}$ and the string mass domain is \cite{2}:  $ 0 \leq  m \leq M_{s}$, ie. $ 0 \leq x \leq 1/2 $,  (which implies for de Sitter background $0\leq \Delta_s\leq  1$,  ie. $1 \leq f(x)\leq 2 $). 
\\
The entropy $S_s(m,H)$ of string states in Anti de Sitter space is larger than the string entropy for $H=0$. The effect of a negative cosmological constant is to increase the entropy, while in de Sitter space the effect of a positive cosmological constant is to reduce the entropy.
For low masses  $m \ll M_{s}$, the entropy is a series expansion in $(H m\alpha' /c)  \ll 1$, that is like a low H expansion around the flat $H=0$ solution, $S_s^{(0)}$ being its leading term. But for intermediate masses  $m \rightarrow M_s$, that is $(H m \alpha^{'}/c) \rightarrow 1$, and for high masses $m >> M_s$ , ie high $ H m \alpha^{'}/c) >>1 $, the situation is very different as we see it below.

Eq.~(\ref{eq:SsH}) is the full string entropy in Anti de Sitter background.  Moreover, Eq.~(\ref{eq:SsHF}) for $S_s(m,H)$ allows us to write in Section~(\ref{sec:dual}) the whole expression for the Anti de Sitter entropy $S_{sem}(H)$, as a function of the Bekenstein-Hawking Anti de Sitter entropy $S_{sem}^{(0)}(H)$.

\section{ High Mass Behavior. Absence of String Branch Point Phase Transition in Anti de Sitter Background\label{sec:sdsps}}

The mass spectrum  and the density of mass levels of quantum string states in $AdS$ background  (Eqs.~(\ref{eq:rhoo}) and~(\ref{eq:rhoc})) do not exhibit explicitely, any upper mass bound as in de Sitter space time. In the de Sitter background, $M_s$ appears as a branch point in the mass spectrum and  mass density of levels.

In Anti de Sitter background, the behavior of the string mass density of levels $\rho_s(m, |\Lambda|)$ for $m\sim M_s$ (Eqs. (\ref{eq:rhoo}) and~(\ref{eq:rhoc})), is
\begin{equation}
\rho_s (m\sim M_s) \sim \Bigg (\frac{M_s}{m_s}\Bigg )^{-a} 
\exp \Bigg\{ \bar{b} ~\frac{M_s} {m_s}\Bigg \}
\label{eq:orhoe} 
\end{equation}

where:\\
$\bar{b} = b  \Big( \frac{\sqrt{5}-1}{2} \Big )^{1/2} $ ~~for open strings, ~~
and ~~$\bar{b} = b  \Big( \frac{\sqrt{2}-1}{2} \Big )^{1/2} $ ~~for closed strings.\\

For $m\gg M_s$, $ \rho (m, |\Lambda|)$ for open strings Eq.~(\ref{eq:rhoo}), behaves as:    

\begin{equation}
\rho_s ~(m \gg M_s) \sim \Bigg (\frac{m}{m_s}\Bigg )^{-(\frac{1+a}{2})} 
\Bigg (\frac{M_s}{m_s}\Bigg )^{(\frac{1-a}{2})} 
\exp \Bigg\{ \frac{b}{m_s} \sqrt{m~M_s} \Bigg\} 
\label{eq:orhomM} 
\end{equation}

and for closed strings, Eq.(\ref{eq:rhoc}):

\begin{equation}
 ~~~~\rho_s ~( m \gg M_s)\sim \Bigg (\frac{m}{m_s}\Bigg )^{-(\frac{1+a}{2})} ~
\Bigg (\frac{M_s}{m_s}\Bigg )^{(\frac{1-a}{2})} \exp \Bigg \{ \frac{b}{m_s} \sqrt{\frac{m~M_s}{2}} \Bigg \}
\label{eq:crhomM} 
\end{equation}
\\
Thus, for intermediate or high masses $m \sim M_s$, and for very high masses  $m\gg M_{s}$, the entropy behaves respectively as:
\begin{equation}
S_s(m \sim M_s) = k_B \left(\bar{b}\frac{M_s}{~m_s}\right)~-~ a k_B \ln \left(\frac{M_s}{m_s}\right)
\label{eq:SsMs}
\end{equation}
\begin{equation}
S_s(m\gg M_{s}) = k_B  \left(\frac{b} {m_s} \sqrt{\frac{~m~M_s}{~2}}\right)~-~ \left(\frac{1 + a}{2}\right)~ k_B \ln~ \left(\frac{m}{m_s}\right)~-~ \left(\frac{1 - a}{2}\right)~ k_B \ln\left(\frac{M_s}{m_s}\right)
\label{eq:SsmMs}
\end{equation}
Or, in terms of temperature : 
\begin{equation}
S_s(T \sim T_s) = k_B  \left(\frac{T_s}{t_s}\right)~- ~ a k_B \ln\left(\frac{T_s}{t_s}\right)
\label{eq:SsTs}
\end{equation}
\begin{equation}
S_s(T\gg T_{s}) = k_B \left(\frac{1} {t_s} \sqrt{\frac{~T~T_s}{~2}}\right)~-~ \left(\frac{1 + a}{2}\right)~ k_B \ln~ \left(\frac{T}{t_s}\right)~-~ \left(\frac{1 - a}{2}\right)~ k_B \ln~ \left(\frac{T_s}{t_s}\right)
\label{eq:SstTs}
\end{equation}

\begin{equation}
T = \frac{1}{ 2 \pi k_B} m c^2.
\end{equation}

We see that for very large $m$, ($m\gg M_{s}$), the leading asymptotic behavior of the $AdS$ string mass density of states grows like $\rho_{s} \sim~ e^{\sqrt{m c/H\hbar}} $ instead of  $e^{m/m_s}$ as in flat space time Eqs.~(\ref{eq:rhofc}) and~(\ref{eq:rhofo}). $\rho _s(m,H)$ expresses in terms of the typical mass scales in each domain: $m/m_s$ (as in flat space) for low masses, $M_s/m_s =(c/H)\sqrt{c/\hbar \alpha'}$ for intermediate masses, $ \sqrt{m M_s}/m_s = c \sqrt{m/\hbar H}$ for very high masses, and there is {\bf no} extra singular factor for high masses as it is the case in de Sitter space. Notice that for the very heavy strings, the mass scale turns out determined by $\hbar H/c^2$ and not by $\sqrt{\hbar/\alpha'c}$. 
\\
The above new features translate into the excited string entropy behavior Eqs.(\ref {eq:SsmMs})-(\ref{eq:SsmMs}) or (\ref{eq:SsTs})-(\ref{eq:SstTs}). We see how the entropy behavior $(m/m_s)$, which in AdS background is a low mass or low curvature behavior $(Hm\alpha'/c <<1)$, does become $M_s/m_s$ and then split into the highly excited entropy behavior $\sqrt{m M_s}/m_s$ in the high mass or high curvature $(Hm\alpha'/c >>1)$ regime. As $Hm$ increases : $m/m_s ~~ \rightarrow ~~  M_s/m_s ~~ \rightarrow ~~ \sqrt{m M_s}/m_s$. 

Furthermore, there is {\bf no} critical string temperature in AdS background, while there exists a critical string temperature in de Sitter background, as well as in the black hole backgrounds. The density of states  and entropy do not show any singular behavior at finite mass or temperature in AdS background. The partition function for a gaz of strings in AdS background is defined for {\bf all} temperature. As a consequence, there is {\bf no} string phase transition of the Hagedorn type in AdS background.  We explicitely show it in Section (\ref {sec:partition}) below. 

For $m \sim M_{s}$, the string is as massive as the background, in other words, the string itself becomes the background, or conversely, the background becomes the string. As a consequence, $M_s$ and its corresponding temperature $T_s$ must be truly considered in practice as limiting values for the string mass and string temperature respectively in AdS background.
 
When the string mass becomes $M_s$,the string size $L_s$ (Compton length for $M_s$) becomes $L_{c\ell}$, the string becomes ``{\it classical} '' reflecting the classical properties of the background, or the background becomes quantum. $M_s$ is the mass of the background $M_{cl}$ Eq.~(\ref{eq:Mcl}), (with $\alpha '$ instead of $G/c^2$): for $m \rightarrow M_ s$ the string becomes  the ``{\it background} '' . Conversely, and interestingly enough, string back reaction supports this fact:  $M_s$ is the mass of  AdS background in its string regime, an Anti de Sitter phase with mass $M_s$ Eq.~(\ref{eq:mMs}) and temperature $T_s$ Eq.~(\ref{eq:Ts}) is sustained by strings . ($L_s$, $M_s$, $T_s$), Eqs.~(\ref{eq:mMs})-(\ref{eq:Ts}), are the {\it intrinsic} size, mass and temperature of $AdS$ background in its string (high H) regime.

\section{Partition Function of Strings in Anti de Sitter Background . Absence of String Critical Temperature in AdS background\label{sec:partition}}
The canonical partition function is given by

\begin{equation}
\ln Z = \frac{V_{D-1}}{(2\pi)^{D-1}} \int^{\infty}_{m_0} d\Big( \frac{m}{m_s}\Big) \rho_s(m, |\Lambda|) ~
\int d^{D-1}k ~\ln \Bigg\{  \frac{1 + \exp \Big\{- \beta_{sem} \Big[(m^2 c^4 + \hbar^2 k^2 c^2)^{1/2}\Big] \Big\}}
{1 - \exp \Big\{- \beta_{sem} \Big[(m^2 c^4 + \hbar^2 k^2 c^2)^{1/2}\Big] \Big\}} \Bigg\} 
\label{eq:Zg}
\end{equation}

where supersymmetry has been considered for the sake of generality; $D-1$ is the number of space dimensions; $\rho_s(m, |\Lambda| )$ is the mass density of states in AdS background; $\beta_{sem}=(k_B T_{sem})^{-1}$ and $T_{sem}$ is the anti de Sitter characteristic semiclassical temperature Eq.~(\ref{eq:Tcl}); $m_{0}$ is the lowest mass for which the asymptotic behavior of $\rho_s(m, |\Lambda|)$ is valid.

From Eq.~(\ref{eq:Zg}),  we have 
\begin{equation}
\ln Z =\frac{4 V_{D-1}}{(2\pi)^{D/2}} \frac{c}{\beta_{sem}^{\frac{D-2}{2}} \hbar^{D-1}}
\sum_{n=1}^{\infty} \frac{1}{(2n-1)^{D/2}} ~
\int^{\infty}_{m_0} d\Big( \frac{m}{m_s}\Big) \rho_s(m, H) m^{D/2}K_{D/2}
\Big[ \beta_{sem} (2n-1)mc^2\Big]
\label{eq:Z}
\end{equation}

Considering the asymptotic behaviour of the Bessel function $K_{\nu}(z)$
\begin{equation}
K_{\nu}(z) \sim \Big( \frac{\pi}{2z}\Big)^2 ~e^{-z}
\label{eq:k}
\end{equation}
and the leading order, $n=1$ $(\beta_{sem} ~m~c^2 \gg 1)$, we have
\begin{equation}
\ln~Z \simeq \frac{2 ~V_{D-1}}{(2\pi )^{\frac{D-1}{2}}}~~ \frac{1}{(\beta_{sem}~~\hbar^2 )^{\frac{D-1}{2}}}~ 
~\int^{\infty}_{m_0} d\Big( \frac{m}{m_s}\Big) ~\rho_s(m, |\Lambda|) ~
m^{\frac{D-1}{2}}~ e^{-\beta_{sem}mc^2}
\label{eq:Zl} 
\end{equation}
(the factor 2 comes from supersymmetry).

$\rho_s(m, |\Lambda|)$ has a different behavior from flat space time and from de Sitter background. It is crucial to notice here that there is not a mass upper bound as in de Sitter case, nevertheless $M_s$ Eq.~(\ref{eq:mMs}) fixes a mass reference in anti de Sitter space time, beyond which AdS string states become highly massive. Let us then analyze $\ln Z$, at significant mass ranges in relation with the string anti de Sitter scale $M_s$. 

For $m \ll M_s$, the $\rho_s(m, |\Lambda|)$ leading behavior is given by the flat  space solution. From Eqs.~(\ref{eq:rhofo}) and (\ref{eq:Zl}), we have for any D-dimensions

\begin{equation}
(\ln Z)_{m\ll M_s} \sim \frac{2 V_{D-1}}{(2\pi)^{ \frac{D-1}{2}}}~
\frac{(m_s)^{\frac{D-3}{2}}}{(\beta_{sem}~\hbar^2)^{\frac{D-1}{2}}}~~
\frac{1}{(\beta_{sem}-\beta_{s})c^2}~e^{-(\beta_{sem}-\beta_{s})~m_0 c^2}
\label{eq:ZmMs}
\end{equation}

\bigskip
where $\beta_{s}=(k_B~t_s)^{-1}$, $t_s$ is given by Eq.~(\ref{eq:ts}).

\bigskip
We see that the canonical partition function, for low $H m \ll  c/\alpha'$, shows a pole singularity at $T_{sem}\rightarrow t_{s}$, Eqs.(\ref{eq:Tcl}) and (\ref{eq:ts}). This transition, near the flat space string temperature $t_s$, and for any space time dimension $D$, is the same as the string transition in flat space time (Carlitz/Hagedorn transition~~\cite{18}). This is also the same as the string transition far from the black hole in the Schwarszchild black hole space-time, and as the string transition in the low curvature regime ($mH \ll c/\alpha'$) of de Sitter space time. For low temperatures $\beta_{sem}\gg \beta_{s}$, (i.e. $T_{sem}\ll t_{s}$),  we recover the non singular Q.F.T exponential decreasing behavior characterized by $T_{sem}$:
\begin{equation}
\ln Z \simeq V_{D-1} ~\Big(\frac {1}{\hbar^2 \beta_s~ \beta_{sem}} \Big)^{\frac{D-1}{2}}~~ e^{- \beta_{sem}~ m_0 c^2 }
\label{eq:ZTh}
\end{equation}

For $m\sim M_s$, $\rho_s(m, |\Lambda|)$ behaves as given by Eq.(\ref{eq:orhoe}) and the leading behavior for the canonical partition function $\ln Z$ is :

\begin{equation}
\ln Z_{m\sim M_s} \sim  \frac {2~V_{D-1}}{(2\pi)^{\frac{D-1}{2}}~ (\hbar c \beta_{sem})^{D-1}}~
 \left(\frac{\beta_s}{\beta_{s~AdS}} \right)^{\frac{3-D}{2}}~e^{\frac{1}{\beta_{s~AdS}}(\beta_s ~-~
  2\pi \beta_{sem})}
 \label{eq: ZmMs}   
\end{equation}

For $m\gg M_s$, the behavior of the string mass density of levels $\rho_s(m, |\Lambda|)$ is given by Eq.(\ref{eq:orhomM}), and the leading behavior for the canonical partition function $\ln Z$ is similar to the $m\sim M_s$ behavior:

\begin{equation}
\ln Z_{m\gg M_s} \sim \frac {2~V_{D-1}}{(2\pi)^{\frac{D-1}{2}} (\hbar \beta_{sem} c)^{D-1}}~
 \left(\frac{\beta_s}{\beta_{s~ AdS}} \right)^{\frac{3-D}{4}}~ ~~e^{\frac{\beta_s}{\beta_s~ AdS}}~~
e^{- 2\pi\frac{\beta_{sem}}{\beta_{s~AdS}}} 
 \label{eq:ZmgMs}
 \end{equation}

where
\begin{equation}
\beta_{s~AdS} = (k_B~ T_s)^{-1}, 
 \end{equation}
~and $T_s$ is given by Eq.(\ref{eq:Ts}).

We see that the canonical partition function $\ln Z$ for excited and highly excited strings in $AdS$ background does {\bf not} feature any singular behavior at $T_s$ in contrast to the string partition function in flat space time which shows a single pole temperature singularity (Carlitz transition), and in contrast to the string partition function in de Sitter space-time which shows a branch point singularity (de Vega-Sanchez transition) ~\cite{15}.  Eqs.(\ref{eq:ZmMs}) , (\ref{eq:ZmgMs}) show that there is {\bf no} string phase transition at $T_s$ for massive and highly massive strings in $AdS$ space-time, that is for $Hm\alpha'/c \sim 1$ and for $Hm\alpha'/c >>1$, in contrast to strings in $dS$ space time.

\section{Schwarzschild black hole - anti de Sitter background\label{sec:SchAdS}}

The D-dimensional Schwarzschild black hole - anti de Sitter space-time ($bhAdS$) is described by the metric

\begin{equation}
ds^{2}=-b(r) ~c^2~dt^{2} + b^{-1}(r)~ dr^2 + r^2 ~d\Omega_{D-2}
 \label{eq:mBAS}
\end{equation}
where 
\begin{equation}
b(r) = 1 -  \frac{r_g}{r}+ \left(\frac{r}{L_{c\ell}}\right)^2,~~~~ r_g =  \Bigg(  \frac{16 \pi ~G~ M}{c^2(D-2) ~A_{D-2}}\Bigg)^{\frac{1}{D-3}},~~~~
A_{D-2} =  \frac{2\pi ^{\frac{(D-1)}{2}}}{\Gamma \Big(\frac{(D-1)}{2}\Big)}
\label{eq:BAS}
\end{equation}

$r_g$ being the Schwarzschild gravitational radius, and $L_{c\ell}$ is given by Eq.~(\ref{eq:Lcl}).

For $D=4$, the equation $b(r)=0$ has one real positive solution, which is the black hole horizon $r_{h}$. The black hole surface gravity is defined by

\begin{equation}
\mathcal{K}_{bhAdS} = \frac{c^2}{2}~ \Bigg|~\frac{db(r)}{dr} ~\Bigg|_{r=r_h}
 \label{eq:KBAS}
\end{equation}
and the black hole QFT temperature  (Hawking temperature) in an $AdS$ background is
given by 
\begin{equation}
T_{sem~bhAdS} = \frac{\hbar}{2~\pi~k_B~c}~\mathcal{K}_{bhAdS} 
\label{eq:TK}
\end{equation}

It is useful to express $T_{sem~bhAdS}$  as

\begin{equation}
T_{sem~bhAdS} = \frac{\hbar~c}{2~\pi~k_B}~~\frac{1}{L_{bhAdS}}
 \label{eq:TL}
\end{equation}

where, for $r_h \simeq r_g$, $L_{bhAdS}$ is given by:
\begin{equation}
L_{bhAdS} = 2~r_g ~ \Bigg(1 + 2~\Big(\frac{r_g }{L_{c\ell}}\Big)^2~~ \Bigg)^{-1}
 \label{eq:L}
\end{equation}

With these expressions for the semiclassical $bhAdS$ background we are prepared to compute the quantum emission of strings by a Schwarzschild black hole in the anti de Sitter background.

\section{Quantum String Emission by a Black Hole in Anti de Sitter Background \label{sec:BHe}}

The quantum field emission cross section $\sigma_{QFT} (k)$ of particles of mode $k$ by a black hole in  Anti de Sitter background is
given by
\begin{equation}
\sigma_{QFT}(k)=\frac{\Gamma_A}{e^{(\beta_{sem~bhAdS} E(k))}-1}
\label{eq:E}
\end{equation}

where $\Gamma_A$ is the black hole greybody factor (absorption cross section), and for the sake of simplicity, only bosonic states have been considered ; $\beta_{sem~ bhAdS}=( k_B T_{sem~bhAdS} )^{-1}$, and $T_{sem~bhAdS}$ is the QFT (Hawking) temperature given by Eq.~(\ref{eq:TL}). The quantum field emission cross section of particles of mass $m$ is defined as

\begin{equation}
\sigma_{QFT}(m) = \int_{0}^{\infty} \sigma_{QFT} (k)~ d\mu (k) 
\label{eq:ED}
\end{equation}

where $d\mu (k)$ is the number of states between $k$ and $k+dk$.

\begin{equation}
d\mu(k) = \frac{2 V_{D-1}}{\Big(4\pi\Big)^{\frac{D-1}{2}}\Gamma \Big( \frac{D-1}{2} \Big) }~k^{D-2}~dk
\label{eq:mu}
\end{equation}

From Eq.~(\ref{eq:ED}) we have
$$ \sigma_{QFT}(m)=\frac{V_{D-1}~\Gamma_A}{(2 \pi)^{\frac{D-1}{2}}}
\frac{\Big( mc^2\Big)^{\frac{D-2}{2}}~}{(\beta_{sem~bhAdS})^{D/2}~ (\hbar c)^{D-1}} ~~
\times $$
\begin{equation}
\sqrt{\frac{2}{\pi}}~\sum_{n=1}^{\infty} \frac{1}{n^{D/2}}~
\Big\{ n\beta_{sem~bhAdS}~ mc^2 K_{_{D/2}} (n\beta_{sem~bhAdS}~mc^2) + K_{_{D/2 - 1}}  (n\beta_{sem~bhAdS}~ mc^2) \Big\} 
\label{eq:smD}
\end{equation}

\bigskip

For large $m$ and the leading order $n=1$, $(\beta_{sem~bhAdS}~ mc^2\gg1)$,  we obtain with the help of Eq.~(\ref{eq:k})

\begin{equation}
\sigma(m)_{QFT} \simeq \frac{V_{D-1}~\Gamma_A} {(2\pi )^\frac{D-1}{2}}~ ~
\frac{m^{\frac{D-1}{2}}} {\left(\beta_{sem~bhAdS} ~\hbar^{2}\right)^ \frac{D-1}{2}}~ e^{-\beta_{sem~bhAdS}~ mc^2}
\label{eq:smDl}
\end{equation}

The quantum string emission, $\sigma_{string}$, is given by
\begin{equation}
\sigma_{string} \simeq  \int_{m_0}^{\infty} \rho_{s}(m, H)~\sigma_{QFT}(m)~ d\Big(\frac{m}{m_s}\Big)  \label{eq:sD}
\end{equation}

where $\rho_{s}(m, |\Lambda|)$ is given by Eqs.~(\ref{eq:rhoo}) and~(\ref{eq:rhoc}).

We consider now the quantum emission for the different mass ranges, with respect to the relevant anti de Sitter string mass scale $M_s$  Eq.~(\ref{eq:mMs}). For low masses $m \ll M_s$, the $\rho_s(m, H)$ leading behavior is given by the flat space solution. From Eqs.~(\ref{eq:smDl}),~(\ref{eq:sD}) and~(\ref{eq:rhoo}) (open strings), the leading contribution to the quantum string emission cross section in any D-dimensions is

\begin{equation}
\sigma_{string}~(m\ll M_s) \sim \frac{V_{D-1}~\Gamma_A} {(2\pi )^\frac{D-1}{2}}~
\frac{ ~m_s^{\frac{D-3}{2}}} {\left(\beta_{sem~  bhAdS}~\hbar^{2}\right)^{\frac{D-1}{2}}}~
\frac{ e^{-(\beta_{sem~bhAdS}~-~\beta_{s})~m_0 c^2}}{\Big(\beta_{sem~bhAdS} -\beta_{s}\Big)c^2} 
\label{eq:smia}
\end{equation}
\\
For $m \ll M_s$, (which is a low regime, $Hm \ll c/\alpha')$ regime, the $bhAdS$ string emission cross section shows the same singular behavior near $t_s$ as the low $Hm$ behavior of the Anti de Sitter partition function Eq.~(\ref{eq:ZmMs}), and as the quantum string emission by an asymptotically flat black hole ~\cite{1},~\cite{4}, in our case here at the bhAdS temperature $T_{sem~bhAdS}$. This is so, since in the bhAdS background, the string mass scale for low string masses (temperatures) is the Hagedorn flat space string temperature $t_s$. The limit $T_{sem~bhAdS} \rightarrow t_s$ is a high temperature behavior for low $Hm \ll c/\alpha'$; $t_s$ is larger than $T_{sem~bhAdS}$ and smaller than the Anti de Sitter string temperature $T_s$.

For low temperatures $\beta_{sem~bhAdS} \gg \beta_{s}$, (i.e. semiclassical regime), we recover the QFT Hawking emission at the temperature  $T_{sem~bhAdS}$:

\begin{equation}
\sigma_{string} (T_{sem~bhAdS} \ll t_{s}) \simeq \frac{V_{D-1}~\Gamma_A}{(2 \pi)^{\frac{D-1}{2}}}~
\frac{m_s^{\frac{D-3}{2}}}{\left(\beta_{sem~bhAdS}\right)~^{\frac{D+1}{2}}~ (\hbar c)^{D-1}}~e^{-\beta_{sem~bhAdS}~ m_0c^2}
\end{equation}
\\
For high masses  $m\sim M_s$, and $m >> M_s$, $\sigma_{string}$ behaves as:

\begin{equation}
\sigma_{string} (m \sim M_s~~\text{and}~~m\gg M_s) \sim \frac{V_{D-1}~\Gamma_A~}{\left(\beta_{sem~bhAdS}~\hbar c\right)^{D-1}}~
\left(\frac{\beta_s}{\beta_{s~AdS}}\right)^{\frac{3-D}{4}}~~
e^{\frac{1}{\beta_{sAdS}}( \beta_s ~-~2\pi \beta_{sem~bhAdS})}~
 \label{eq:sigma3}
\end{equation}

The evaporation of a black hole in $AdS$ space time, from its early  semiclassical or quantum field theory regime (Hawking radiation) towards a quantum string anti de Sitter regime (late stages), can be seen as well from the black hole decay rate. In the early evaporation stages, the  semiclassical black hole in anti-de Sitter background decays thermally as a grey body at the Hawking temperature $T_{sem~bhAdS}$ Eq.~(\ref{eq:TL}), with the decay rate

\begin{equation}
\Gamma_{sem} = \left| \frac{ d\ln M_{sem~bhAdS} }{ d t}\right| \sim G~\left(T_{sem  ~bhAdS}\right)^{3} , ~~~~~~~~M_{sem~bhAdS}= 2~\pi~T_{sem~bhAdS}    \label{eq:decay}
\end{equation}

($\hbar=c=k_{B}=1$).\\ 
As evaporation proceeds, $T_{sem~bhAdS}$  increases until it reaches the string anti de Sitter temperature  $T_{s}$ Eq.~(\ref{eq:Ts}), the black hole itself becomes an excited string anti de Sitter state, decaying with a string width $\Gamma_{s}\sim \alpha'~T_{s}^{3}$ ,~~$(G\sim \alpha')$,  into all kind of particles, with pure (non mixed) quantum radiation. For $T_{sem~bhAdS} \rightarrow T_s$, the semiclassical black hole decay rate $\Gamma_{sem}$ becomes precisely the quantum string decay $ \Gamma_s $. The implications of the limit $T_{sem~bhAdS} = T_s$ are analyzed in the Section below.

\section{String bounds for a black hole in Anti de Sitter background\label{sec:bBHS}}

Far from the black hole, the black hole-Anti de Sitter ($bhAdS$) background tends asymptotically to $AdS$ space-time. Black hole evaporation will be measured by an observer which is at this asymptotic region. Asymptotically, in the Schwarzschild black hole-Anti de Sitter space-time, $\rho_s(m,H)$ is equal to the string mass density of states in $AdS$ space time Eq.~(\ref{eq:rhoo}). Then, for the partition function of a gaz of strings far from the black hole in the $bhAdS$ space-time, we only need to substitute  $\beta_{sem}$  in   Section (\ref{sec:partition}) by $\beta_{sem~bhAdS}$,  i.e.,  substitute the Anti de Sitter temperature $T_{sem}$ Eq.~(\ref{eq:Tcl}) by the black hole temperature in Anti de Sitter space $T_{sem~bhAdS}$ Eq.~(\ref{eq:TL}). With this substitution,  all results in Section (\ref{sec:partition}) hold for the $bhAdS$ case as well. Mathematically, there is no strict bound $T_{sem~bhAdS} < T_s$ emerging from the string partition function in the bhAdS case, since the bhAdS string partition function is mathematically well defined for all temperature. However, the regime when $T_{sem~bhAdS}$ reaches the string temperature $T_s$  truly characterizes the string regime of the $bhAdS$ background. From Eqs.~(\ref{eq:Tcl}),~(\ref{eq:Ts}),~(\ref{eq:TL}),~(\ref{eq:L}), the condition  
\begin{equation}
T_{sem~bhAdS} = T_s,  
\label{eq:TsemTs}
\end{equation}
yields :
\begin{equation}
\ell_s^2 = L_{c\ell}~L_{bhAdS} ~~, 
\label{eq:lLbhAdS}
\end{equation}
which implies the following equation 
\begin{equation}
\ell_s^2~\Big[1 + 2\left(\frac{r_g}{L_{c\ell}}\right)^2\Big] = 2~r_{g} L_{c\ell}
\label{eq:HbS}
\end{equation}

Thus, the black hole gravitational radius $r_g$ satisfies 
\begin{equation}
\left(\frac{r_g}{L_{c\ell}}\right)^{2}~-~r_g~ \frac{L_{c\ell}}{\ell_s^2}~+~\frac{1}{2} = 0
\label{eq:rgb}
\end{equation}
with the solution:

\begin{equation}
r_{g \pm}~  = ~\frac{1}{2}~\frac{L_{c\ell}^3}{\ell_s^2}~\Big[ ~ 1~ \pm~\sqrt{ 1 - 2\left(\frac{\ell_s}{L_{c\ell}}\right)^4~}~\Big]
\label{eq:rgs}
\end{equation}

Both $r_{g +}$ and $r_{g -}$ are physical roots provided:
\begin{equation}
\label{eq:rgc}
L_{c\ell}~ \geq 2^{1/4}~\ell_s~\equiv L_{c\ell~min}~=~ 1.189~ \ell_s
\end{equation}

That is, there is a {\it minimal} AdS classical length $L_{c\ell~min}$, or a {\it maximal} AdS string length $L_{s~max}$ :  
\begin{equation}
L_s~\leq~ 2^{-1/4} \ell_s~\equiv ~L_{s~max} ~=~ 0.841~\ell_s.
\label{eq:rgams}
\end{equation}

For $L_{c\ell} \gg L_{c\ell~min}$, $r_{g +}$ and $r_{g -}$ are : 
\begin{equation}
r_{g +} ~ =~ \frac {L_{c\ell}^3} {\ell_s^2}~\Big[1 - \frac{1}{2}\left(\frac {\ell_s}{L_{c\ell}}\right)^4 + O\left(\frac {\ell_s}{L_{c\ell}}\right)^8 ~\Big] 
\label{eq:rga}
\end{equation}
\begin{equation}
r_{g -} ~=~ \frac {1}{2}~\frac{\ell_s^2} {L_{c\ell}}~\Big[~1 - O\left(\frac {\ell_s}{L_{c\ell}}\right)^2 ~\Big] ~~,
\label{eq:rgg}
\end{equation}
which in terms of $L_s$  read: 
\begin{equation}
r_{g +} ~ \simeq \frac {L_{c\ell}^2}{L_s}=\frac {c^4}{\hbar\alpha'H^3}~~~~,~~~~r_{g -} \simeq \frac {1}{2}~ L_s= \frac {\hbar\alpha'H}{2c^2}~~~~,~~~{\text ie}~~~  ~r_{g +}r_{g-}= \frac{L_{c\ell}^2}{2} 
\label{eq:rgam}
\end{equation}

On the other hand, for  $L_{c\ell}~ =~ L_{c\ell~min}~=~ 2^{1/4}~~\ell_s$, (or $L_s~ =~ L_{s~max}~=~ 2^{-1/4}~~\ell_s$),
$r_{g~ \pm}$ are:
\begin{equation}
r_{g +}~ = ~ r_{g -}~ = ~ \frac {L_{c\ell~min}}{\sqrt{2}}= L_{s~max} =  \frac { \ell_s}{2^{1/4}}= 0.841~ \ell_s  
\label{eq:rgamss}
\end{equation}
\\
Eq.~(\ref{eq:rgs}) shows the relation between the  Schwarzschild radius and the cosmological constant Eq.~(\ref{eq:Lcl}) when $T_{sem~bhAdS} = T_s$ ( bhAdS string regime). We see that a black hole in Anti de Sitter space allows an intermediate string regime, not present in the Schwarzschild black hole alone $(H=0)$, since in the bhAdS background there are two characteristic string scales: $ L_s$ and $\ell_s$. In an asymptotically flat space-time, the black hole radius becomes $\ell_s$ in the string regime. In an asymptotically Anti de Sitter space, when $T_{sem~bhAdS}$ reaches $T_s$ , the black hole radius $r_g$ becomes either $r_{g +}$ or $r_{g -}$, depending on the Anti de Sitter string size $L_s$. Then, when the AdS characteristic length $L_{c\ell}$ reaches its minimal value, $r_g$ becomes uniquely determined by $\ell_s$, as given by Eq.(\ref{eq:rgamss}). The bhAdS string regime determines a maximal value for $H$:  $H_{max} = 0.841~c/\ell_s$

The bhAdS string regime is larger than black hole de Sitter (bhdS) string regime. Notice the differences between the bhAdS and bhdS string regimes: \\
(i) only one root ($r_{g-}$) is present in the bhdS string regime.\\ 
(ii) there is no condition such as Eq.(\ref{eq:rgc}) or (\ref{eq:rgams}) for $L_{c\ell}$ or $L_s$ in the bhdS string regime. \\
(iii) 
In the bhdS space, when $T_{sem~bhdS}$ reaches $T_s$, the black hole radius $r_g$ becomes $L_s$. Then, when $L_{c\ell} = \ell_s$, $r_g$ is minimal and determined by $\ell_s$ too,  ($r_{g~min} = 0.365~\ell_s$).\\ 
In contrast, in the bhAdS background, $L_{c\ell}$ can not reach $\ell_s$, $L_{c\ell~min}$ is  larger than $\ell_s$, Eq.(\ref{eq:rgc}). The minimal black hole radius in AdS space-time, $r_{g~ min} = 0.841~  \ell_s$, is {\bf larger} than the minimal black hole radius in de Sitter space: 
\begin{equation}
r_{g~min~bhAdS} = 2.304~ r_{g~min~bhdS}
\end{equation}

\section{Semiclassical (Q.F.T) and quantum (string) Anti de Sitter regimes\label{sec:dual}}

We have shown that for  $m\rightarrow M_s$,~ i.e. $T\rightarrow T_s$. From the string canonical partition function Sec.~(\ref{sec:partition}) in $AdS$ space time, the microscopic string density of mass states $\rho_s(m, H)$ Secs. (\ref{sec:qs}) and (\ref{sec:sdsps}), the quantum string emission by a black hole in $AdS$ space time Sec.~(\ref{sec:BHe}) and the bhAdS string bounds Sec.~(\ref{sec:bBHS}), we have shown that, for $T_{sem}\rightarrow T_s$, the semiclassical (Q.F.T) regime becomes a string regime at the string Anti de Sitter temperature $T_{s}$ Eq.~(\ref{eq:Ts}). This means that, in the quantum string regime, $L_{c\ell}$ becomes $L_{s}$ , $T_{sem}$ becomes $T_s$, the semiclassical density of states $\rho_{sem}$ and semiclassical entropy $S_{sem}$ become the string density  of states $\rho_s$ and entropy $S_{s}$. Namely, a semiclassical Anti de Sitter state, $(AdS)_{sem}= (L_{c\ell}, M_{c\ell},  T_{sem}, \rho_{sem}, S_{sem})$, becomes a quantum string state $(AdS)_{s}$ = $(L_{s}, m , T_{s}, \rho_{s}, S_{s})$.\\ 
The sets $(AdS)_{s}$ and $(AdS)_{sem}$ are the same quantities but in different (quantum and semiclassical) regimes. This is the usual classical/quantum duality but in the gravity domain, which is {\it universal}, not linked to any symmetry or isommetry nor to the number or the kind of dimensions. 
From the semiclassical and quantum Anti de Sitter regimes $(AdS)_{sem}$ and $(AdS)_{s}$, we can write the full Anti de Sitter entropy $S_{sem}(H)$, with quantum corrections included, such that it becomes the Anti de Sitter string entropy $S_s(m,H)$ Eq.~(43) in the string regime. Therefore,  the full Anti de Sitter entropy $S_{sem}(H)$ is given by

\begin{equation}
S_{sem}~(H) = \hat{S}_{sem}^{(0)}~(H)
-a~k_B~\ln ~(\frac{\hat{S}_{sem}^{(0)}~(H)}{k_B}) - k_B \ln~F(H)
\label{eq:SsemHF}
\end{equation}
where
\begin{equation}
\hat{S}_{sem}^{(0)}~(H)\equiv S_{sem}^{(0)}~(H) \sqrt{f(X)}~~~~,~~~~F(H)\equiv \sqrt{(1 + 4X^2)f(X)}    
\label{eq:Fsem}
\end{equation}
 
\begin{equation}
a=D~~ ,~~~~~f(X)= \frac{2}{1+\Delta},~~~~ \Delta \equiv\sqrt{1 + 4X^2}~=~ \sqrt{1 +\Big(\frac{\pi k_B}{S_{sem}^{(0)}(H)}\Big)^2} 
\label{eq:Delta}
\end{equation}
 
\begin{equation}
2 X(H)\equiv \frac{\pi k_B}{S_{sem}^{(0)}(H)}= \frac {M_{sem}}{M_{cl}}= \Big(\frac{m_{Pl}}{M_{cl}}\Big)^2   
\label{eq:X}
\end{equation}
\\
$S_{sem}^{(0)}(H)$ is the Bekenstein-Hawking entropy for $AdS$ space time Eq. (9). $M_{cl}$ is the Anti de Sitter mass scale Eq.(\ref{eq:Mcl}), $M_{sem}$ is the semiclassical mass Eq.(\ref{eq:Msem}). In terms of $S_{sem}^{(0)}(H)$, $S_{sem}(m,H)$ Eq.(\ref{eq:SsemHF}) reads:

\begin{equation}
S_{sem}(H) = \sqrt{f(X)} S_{sem}^{(0)}(H) 
-ak_B~\ln \Big( \sqrt{f(X)} \frac{S_{sem}^{(0)}(H)}{k_B} \Big) - k_B~\ln\sqrt{ f(X)}- k_B~\ln\sqrt{~1 + 4X^2~}
\label{eq:SsemH}
\end{equation}

$\Delta$  Eq.(\ref{eq:Delta}) with $X(H)$ Eq.(\ref{eq:X}) describes $S_{sem}(H)$ in the mass domain  $ 0 \leq M_{c\ell} \leq \infty$, that is,  $0 \leq X \leq \infty$,  ~$1\leq \Delta \leq \infty$,~ $ 0 \leq f(X) \leq 1$. 

\bigskip

Eq.~(\ref{eq:SsemHF}) provides the whole Anti de Sitter entropy $S_{sem}(H)$ as a function of the Bekenstein-Hawking entropy $S_{sem}^{(0)}(H)$. The limit $X \rightarrow 0$ means $M_{cl}\gg m_{Pl}$,  that is  $L_{c\ell}\gg \ell_{Pl}$ , (low $H \ll c/\ell_{Pl}$ or low curvature regime). In this classical limit:  $\Delta \rightarrow 1$,  $f(X)\rightarrow 1$ and  $S_{sem}^{(0)}(H)$  is the leading term of $S_{sem}(H)$, with its logarithmic correction:
\begin{equation}
S_{sem}(H) = S_{sem}^{(0)}(H) -ak_B~\ln \Big(\frac{S_{sem}^{(0)}(H)}{k_B}\Big) 
\label{eq:SsemoH}
\end{equation}

The classical AdS regime is like the de Sitter or the Schwarschild black hole classical regime  ($M_{cl}\gg m_{Pl}$): in this regime, the leading term to the entropy is the Bekenstein-Hawking entropy.

But for {\bf high} Hubble constant, $H \sim c/\ell_{Pl}$, (ie. $M_{cl}\sim m_{Pl}$), the high curvature or quantum AdS regime is very different from the high curvature or quantum de Sitter regime, as we precisely see below. 

\section{ Absence of Anti de Sitter Gravitational Phase Transition \label{subsec:EXT}}

In terms of the mass, or temperature, the variable $\Delta$ reads:
\begin{equation} 
\Delta~ =~ \sqrt{1 +  \Big(\frac{m_{Pl}}{M_{cl}} \Big)^4}~=~ \sqrt{1 + \Big(\frac{T_{sem}}{T} \Big)^2} ~~,
\label{eq:Deltasem}
\end{equation}
where
\begin{equation}
T = \frac{1}{2\pi k_B}M_{cl} c^2
\label{eq:T}
\end{equation}

and $T_{sem}$ is the semiclassical Anti de Sitter temperature Eq.(\ref{eq:Tcl}) or Eq.(\ref{eq:TM}).

It must be noticed that $\Delta$ {\it never} vanishes in AdS space-time, contrary to de Sitter space-time in which $\Delta \rightarrow 0 $ for $ M_{c\ell}\rightarrow M_{sem}$, that is for  $M_{c\ell}\rightarrow m_{Pl}$.

For $M_{c\ell}\rightarrow m_{Pl}$, the Bekenstein-Hawking AdS entropy is : $S_{sem}^{(0)}(H)(M_{c\ell}= m_{Pl})= \pi k_B$, and the full  AdS entropy $S_{sem}(H)$ Eq.~(\ref{eq:SsemHF}) is :
\begin{equation}
S_{sem}(H) = \sqrt{\frac{2}{1+\sqrt{2}}}~ \pi k_B - ak_B~\ln \Big(\sqrt{\frac{2}{1+\sqrt{2}}}~\pi \Big) - k_B~\ln \Big(\frac{2}{\sqrt{1+\sqrt{2}}} \Big) 
\label{eq:SsemMcl}
\end{equation}

\bigskip

Contrary to de Sitter space,  {\bf no} phase transition occurs at $T \rightarrow T_{sem}$ in de Anti de Sitter space.  That is, there is {\bf no} phase transition at $M_{cl} \rightarrow m_{Pl}$, ie $T \rightarrow t_{Pl}$, (high curvature $H \rightarrow c/\ell_{Pl}$ or quantum AdS regime).  This is so, since for AdS background,
like for the AdS string entropy, no singularity at finite mass or finite temperature occurs in the density of states or entropy.
Also contrary to de Sitter space, and to the rotating black hole (extremal transition), there is not a square root branch point in the mass (temperature) analogous to the thermal self-gravitating gas phase transition of point particles~~\cite{15}.

\bigskip

It must be noticed that in $AdS$ background we can still analyze the mass regime  $M_{cl} << m_{Pl}$, for  which $ X >> 1$, $ \Delta = 2X >> 1$ and $ f(X) = 1/X  \rightarrow 0 $. In this regime, the Bekenstein-Hawking term is subdominant:
\begin{equation}
S_{sem}^{(0)}(H)(X >> 1)= \frac {\pi k_B}{2X} << 1, 
\label{eq:Ssempl}
\end{equation}

therefore,   $\sqrt {f(X)}S_{sem}^{(0)} = \pi k_B /( 2 X^{3/2})$ , and the full AdS entropy $S_{sem}(H)$ Eq.~(\ref{eq:SsemHF}) beheaves as :
\begin{equation}
S_{sem}(H) (X >>1) = ak_B~\ln \Big(\frac{2}{\pi}~ X ^{3/2}\Big) - k_B~ \ln \Big (2X^{1/2}\Big) ~+~ \frac{\pi k_B}{2X^{3/2}}
\label{eq:SsmMmpl}
\end{equation}

This is a {\bf high} curvature quantum regime, in which $H >> c/\ell_{Pl}$, and therefore, the entropy is not dominated by the usual Bekenstein-Hawking (zero order term), which results negligable in this case. The Planck scale  $Ads$ curvature regime ($H \sim c/\ell_{Pl}$, is reached {\bf smoothly}, without any any singular behavior or phase transition in the entropy. The very high curvature $AdS $ regime $H >> c/\ell_{Pl}$, is reached with a logarithmic growing behavior of the entropy. 

\begin{acknowledgements}
A. B. acknowledges the Observatoire de Paris, LERMA, for the kind hospitality extended to him.
M. R. M. acknowledges the Spanish Ministry of Education and Science ( FPA 2004-2602 project) for financial
support, and the Observatoire de Paris, LERMA, for the kind hospitality extended to her.
\end{acknowledgements}


\end{document}